\documentclass[11pt]{article}
\usepackage{blois,epsfig}

\def\be{\begin{equation}}
\def\ee{\end{equation}}
\def\bea{\begin{eqnarray}}
\def\eea{\end{eqnarray}}

\begin{document}
\vspace*{4cm}
\title{EARLY QUINTESSENCE AND THE CMB}

\author{ CHRISTIAN M.\ MUELLER }

\address{Institut f\"{u}r Theoretische Physik, Philosophenweg 16 \\
69120 Heidelberg, Germany}

\newcommand{\omegals}{\Omega^{(\rm ls)}_{\rm q}}
\newcommand{\omegasf}{\Omega^{(\rm sf)}_{\rm q}}
\newcommand{\wbar}{\overline{w}_{\rm q}}
\newcommand{\w}{w_{\rm q}}
\newcommand{\omegaq}{\Omega_{\rm q}}


\maketitle\abstracts{
We examine the cosmic microwave background (CMB) anisotropy for signatures of early
quintessence dark energy -- a non-negligible quintessence energy density during the
recombination and structure formation eras. In contrast to a 
$\Lambda$CDM cosmology, early quintessence leads to a larger suppression of structure growth. 
Because of this influence on the clustering of dark matter
and the baryon-photon fluid, we may expect to find trace signals in the CMB and the
mass fluctuation power spectrum. In detail, we demonstrate that suppressed
clustering power on small length-scales, as suggested by the combined Wilkinson
Microwave Anisotropy Probe (WMAP) / CMB / large scale structure data set, is
characteristic of early quintessence. }


\section{Introduction}
There exists compelling evidence that the energy density of the Universe is
dominated by dark energy as demonstrated
by the recent high precision measurement of the cosmic microwave background (CMB)
fluctuations by WMAP.
\cite{Bennett:2003bz,Spergel:2003cb,Kogut:2003et,Hinshaw:2003ex,Verde:2003ey,Page:2003fa}
And yet, the nature of the dark energy remains elusive.  A cosmological constant
($\Lambda$), providing a simple phenomenological fix in the absence of better
information, is consistent with current data including the latest WMAP results. 
Lessons from particle physics and cosmology, however, suggest a more attractive
solution in the form of a dynamical dark energy
\cite{Wetterich:fm,Ratra:1987rm,Peebles:1987ek,Caldwell:1997ii} that continues to
evolve in the present epoch  --- quintessence. 

In the basic quintessence scenario, the dark energy enters only at late times, as
required for cosmic acceleration. In a more realistic picture, the late appearance
of the quintessence may not be the whole story.  Scalar field models of quintessence
with global attractor solutions
\cite{Wetterich:fm,Ratra:1987rm,Steinhardt:nw} have been shown to
``track'' the dominant component of the cosmological fluid. One consequence is that
just after inflation, the universe may contain a non-negligible fraction of the
cosmic energy density. Through subsequent epochs, the quintessence energy density
$\rho_{\rm{q}}$ lags behind the dominant component of the cosmological fluid with a
slowly varying $\omegaq$, and an equation-of-state $\w \equiv
p_{\rm{q}}/\rho_{\rm{q}}$ which is nearly constant.  The field energy tracks the
background until the current epoch, when the quintessence energy density crosses and
overtakes the matter density. A non-negligible fraction of dark energy at last
scattering, $\omegals$, and during structure formation, $\omegasf$, then arises
quite  naturally. From the observational viewpoint, detection of any trace of
``early quintessence'' would give us a tremendous clue as to the physics of dark
energy. For an extended discussion on ``early quintessence'' 
and its relation to the WMAP data, see Caldwell 
et al.\cite{Caldwell:2003vp}

In this work we concentrate on ``early quintessence'', characterized by
non-negligible values $\omegals,\, \omegasf \stackrel{<}{\sim } 0.05$. Typical scalar field
models exhibit an exponential form of the scalar potential in the range of the field
relevant for early cosmology,  with special features in the potential or kinetic
term in the range governing the present epoch.
\cite{AW,Albrecht,Armendariz,Wetterich:2003jt} Our attention is drawn
toward these models due to the recent claims of suppressed power on small scales in
the combined WMAP / CMB / large scale structure data set. We are motivated precisely
by the fact that the most prominent influence of a small amount of early dark energy
is a suppression of the growth of dark matter fluctuations. \cite{Doran:2001rw,FJ} 
As we soon discuss, this influence can help to make the fluctuation amplitude
extracted from galaxy catalogues or the Ly-$\alpha$ forest compatible with a
relatively high amplitude CMB anisotropy.  
The effect of early quintessence on the mass fluctuation power spectrum can be
understood simply as a suppression of the growth function for dark matter and
baryonic fluctuations. Just as fluctuation growth is suppressed at late times with
the onset of dark energy domination, so is the growth of linear modes slowed at early times due
to non-negligible $\omegals,\,\omegasf$. We emphasize that this
results in a uniform suppression of the cold dark matter
amplitudes for all modes that have entered the horizon since $z_{eq}$.

\begin{figure}[ht]
\begin{center}
\epsfig{figure=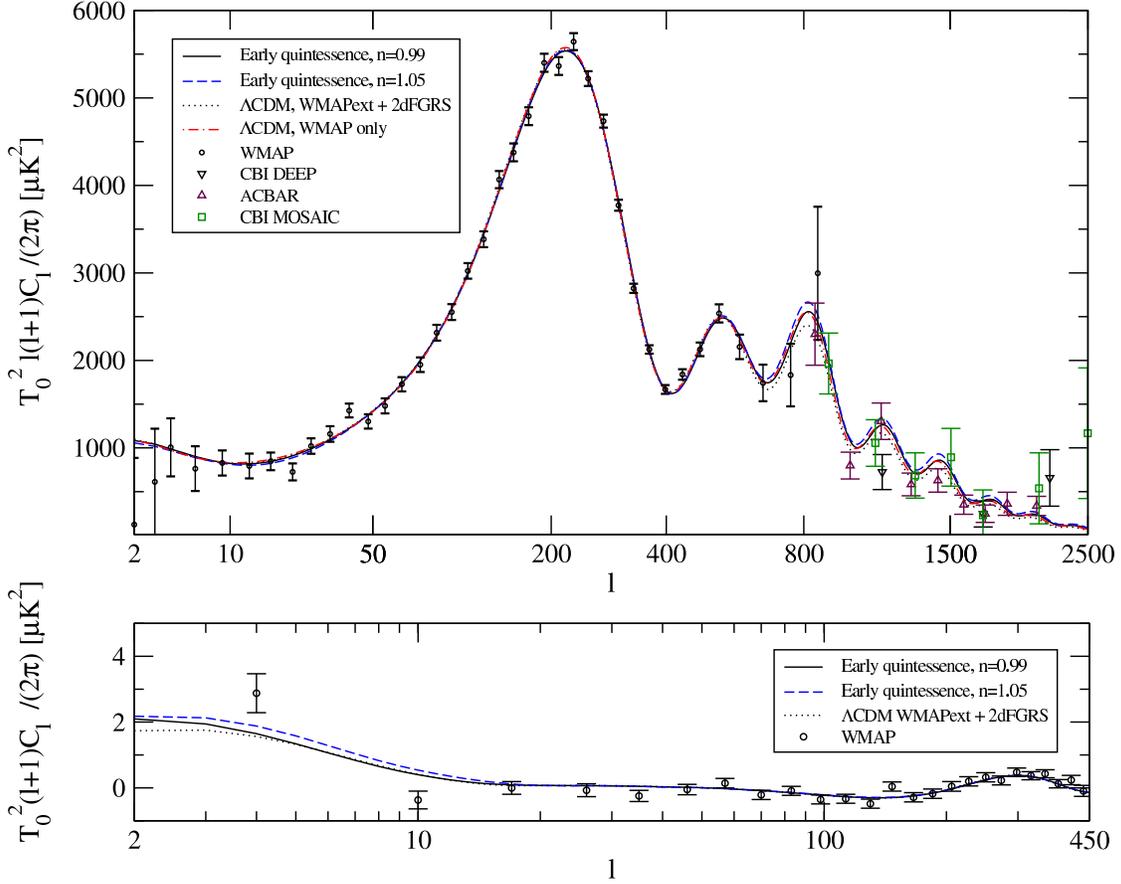,height=5in}
\caption{Temperature (TT) and Polarization (TE) as a function of multipole $l$. The WMAP data 
        \protect\cite{Kogut:2003et,Hinshaw:2003ex} are plotted alongside
        two early quintessence models with $n_s=0.99$ and $n_s=1.05$ (see
        Table \ref{models} for the other cosmological parameters). For comparison,
        we plot WMAP-normalized spectra for the best fit $\Lambda$CDM model (no 
        Ly-$\alpha$ data) with constant spectral index $n=0.97$ of Spergel et al.,
        \protect\cite{Spergel:2003cb} as well as the best fit $\Lambda$CDM model with
        running spectral index $n_s=0.93,\ {\rm d} n_s / {\rm d} \ln k = -0.031$.
        At large $l$ we plot the CBI \protect\cite{CBI_mosaic,CBI_deep} and ACBAR
        \protect\cite{ACBAR} measurements.
\label{cmb}}
\end{center}
\end{figure}

Now we turn to consider the implications of the CMB for quintessence.\cite{DLSW} For the case of
quintessence, a degeneracy exists among the cosmological parameters,\cite{Caldwell:2003vp} 
for instance the equation-of-state can play off the Hubble constant to achieve an otherwise
indistinguishable anisotropy pattern out to small angular scales. \cite{Huey:1998se}
As a means of proof by example, we identify a set
of models in Table~\ref{models} with observationally indistinguishable CMB patterns,
{\it i.e.} identical peak positions, but differing amounts of $\omegals,\,\omegasf$,
shown as Models (A,B) in Figure~\ref{cmb}. Model (C) is WMAP's best fit $\Lambda$CDM
and Model (D) the best fit for an extended data set with $\Lambda$CDM and running
spectral index. \cite{Spergel:2003cb} Clearly, the CMB sky is consistent with a small amount of early quintessence in
addition to $\omegaq^{(0)}$ insofar as the angular-diameter distance to the last
scattering surface is preserved.
For a more thorough treatment of the concordance regions see 
Caldwell \& Doran.\cite{Caldwell:2003hz}

\begin{table}[ht]
\vspace{0.4cm}
\caption{Cosmological parameters of the models presented in this work.\label{models}}
\begin{center}
\begin{tabular}{|c|c|c|c|c|}
\hline
                 & A & B & C & D  \\ \hline 
  $\omegasf$     &  0.03 & 0.05  & 0  & 0  \\
  $\omegals$     &  0.03 &  0.05 & 0  & 0  \\ 
  $\w^{(0)}$     & -0.91 & -0.95  & -1   & -1      \\ 
  $n_s$          & 0.99  & 1.05  & 0.97   & 0.93 \\ 
  $h$            & 0.65  & 0.70  & 0.68   & 0.71  \\ 
  $\Omega_m h^2$ & 0.15  & 0.16  & 0.15  & 0.135 \\
  $\Omega_b h^2$ & 0.024 & 0.025  & 0.023 & 0.0224 \\
  $\tau$         & 0.17  & 0.26   & 0.1     & 0.17    \\  \hline 
  $\sigma_{8}$   & 0.81  & 0.87 & 0.87 & 0.85 \\
  $\chi^2_{eff} / \nu$ &1432/1342  & 1432/1342  & 1430/1342 &  1432/1342\\

\hline
\end{tabular}
\end{center}
\end{table}


We have computed the spectra in Figure~\ref{cmb} using
\mbox{CMBEASY} \cite{Cmbeasy} for a class of ``leaping kinetic term quintessence''
\cite{AW} models that feature early quintessence. The main features depend only on
two parameters besides the present fraction of dark energy $\omegaq^{(0)}$ and the
present equation of state $\w^{(0)}$, namely the fraction of dark energy at last
scattering, $\omegals$, and during structure formation, $\omegasf$. In  order to
facilitate comparison with other effects of quintessence -- for example the Hubble
diagram $H(z)$ for supernovae -- we present a useful parametrization of quintessence
\cite{Wetterich:2003jt} rather than detailed models. For $a> a_{eq}$ and $x \equiv \ln a = - \ln
(1+z)$ we consider a quadratic approximation for the averaged equation-of-state
$(x_{ls} \approx -\ln(1100))$
\begin{equation}
\wbar(x) = - \frac{1}{x}\int^0_x dx' \w(x') = \w^{(0)} + (\wbar^{(ls)}-\w^{(0)}) \frac{x}{x_{ls}} + A x (x-x_{ls}).
\end{equation}
The time-dependent average equation of state $\wbar(x)$ is directly connected to the
time history of the fraction in dark energy $\omegaq (x)$ as one may verify.
 The parameter $A$ is
related to the average fraction of dark energy during structure formation. The parameters describing our models are  (A): $\wbar^{(\rm ls)}=-0.188$,
$A=-0.0091$; (B): $\wbar^{(\rm ls)}=-0.172$, $A=-0.015$. 
Our models (A) and (B) are consistent with the large scale structure observations of the 2dFGRS survey \cite{Percival,Verde,Peacock} and all present bounds for $H(z)$, including
type Ia supernovae.
\cite{Schmidt:1998ys,Riess:1998cb,Garnavich:1998th,Perlmutter:1998np,Perlmutter:1999jt}


To summarize, we have demonstrated that models of early quintessence are compatible
with the presently available data for a constant spectral index of primordial
density perturbations. The presence of early quintessence results in a reduction in the spectrum of
matter fluctuations on small scales, which may have significant consequences for the
interpretation of combined CMB and large scale structure data.

\section*{Acknowledgements}
We thank Robert R. Caldwell, Michael Doran, Gregor Sch{\"{a}}fer and Christof Wetterich for 
helpful discussions.  C.M. M{\"{u}}ller is  supported by GRK grant 216/3-02.


\end{document}